\documentclass[amsmath,amssymb,amsbsy,prl,nofootinbib,twocolumn,tightenlines,superscriptaddress,floatfix]{revtex4-2}
\usepackage{amsfonts} 
\usepackage{amsmath} 
\usepackage{amssymb}
\usepackage{blindtext}
\usepackage{lineno}
\usepackage{cancel}
\usepackage{bm}
\usepackage{graphicx,nicefrac}
\usepackage{xfrac}
\usepackage[usenames, dvipsnames]{color} 
\usepackage{dcolumn}
\usepackage{epic} 
\usepackage{epsfig}
\usepackage{eucal} 
\usepackage{esdiff}
\usepackage{graphicx}
\usepackage[breaklinks,colorlinks = true,linkcolor = blue,urlcolor=blue,citecolor=red]{hyperref}

\begin{document}

\title{Parity-Dependent Scaling of Velocity-Gradient Correlations in Turbulence}
\author{Anwesha Dey$^\clubsuit$}
\email{anwesha.dey@icts.res.in}
\affiliation{International Centre for Theoretical Sciences, Tata Institute of Fundamental Research, Bengaluru 560089, India}
\author{Ritwik Mukherjee$^\clubsuit$}
\email{ritwik.mukherjee@icts.res.in}
\affiliation{International Centre for Theoretical Sciences, Tata Institute of Fundamental Research, Bengaluru 560089, India}
\author{Aikya Banerjee}
\email{aikya.banerjee@physics.ox.ac.uk}
\affiliation{Department of Physics, University of Oxford, Denys Wilkinson Building, Keble Road, Oxford OX1 3RH, United Kingdom}
\affiliation{Merton College, Merton Street, Oxford, OX1 4JD, United Kingdom}
\author{Samriddhi Sankar Ray}
\email{samriddhisankarray@gmail.com}
\affiliation{International Centre for Theoretical Sciences, Tata Institute of Fundamental Research, Bengaluru 560089, India}

\begin{abstract}

We investigate two-point velocity-gradient correlation functions in homogeneous
	isotropic turbulence using exact relations and direct numerical
	simulations. The second-order gradient correlation is shown to be
	exactly related to the Laplacian of the velocity correlation, implying
	inertial-range scaling $C_2^{1,1}(r)\sim r^{-4/3}$. At higher orders,
	we uncover a parity-dependent organization of gradient correlations:
	odd--odd correlations exhibit scaling close to $r^{-4/3}$ with weak
	dependence on order, whereas even--even correlations display
	systematically different exponents. We show that this distinction
	originates from the sign structure of the gradient field: sign
	decorrelation suppresses intermittent contributions in odd--odd
	sectors, while even--even correlations retain them and remain sensitive
	to the spatial organization of intense structures. The measured
	even--even exponents are quantitatively consistent, across two Reynolds
	numbers, with independently measured box-counting dimensions of
	intermittent gradient structures. These results identify parity under
	sign reversal as a fundamental organizing principle for higher-order
	turbulent correlations and establish a direct connection between sparse
	intermittent geometry and scaling exponents in turbulence.

\end{abstract}

\maketitle

\def\thefootnote{$\clubsuit$}\footnotetext{These authors contributed equally to this work}\def\thefootnote{\arabic{footnote}}


The inertial range of fully developed, homogeneous and isotropic turbulence,
$\eta \ll r \ll L$, exhibits robust scaling laws across experiments and
simulations~\cite{Frisch-book}. Kolmogorov phenomenology predicts
inertial-range scaling behaviour such as $E(k)\sim k^{-5/3}$ and $S_p(r)\sim
r^{p/3}$~\cite{kolmogorov1941}, modified by intermittency.  However, these
ideas have been developed primarily for velocity increments and related
observables. In contrast, the most intense fluctuations occur in velocity
gradients $\partial_j u_i$, which govern dissipation, small-scale dynamics, and
chaotic
stretching~\cite{JohnsonWilczekReview2024,DhawalPumir2026,Murugan2021,Banerjee2026}, yet
their spatial correlations remain largely unexplored.

Single-point velocity-gradient statistics are strongly intermittent and
non-Gaussian, reflecting rare, intense
events~\cite{Nelkin1990,SreenivasanAntonia1997,Schumacher2007,meneveau2011,Chakraborty2012,JohnsonWilczekReview2024,DhawalPumir2026}.
However, how these features manifest in two-point correlations is not
understood. In particular, it remains unclear whether velocity-gradient
correlations exhibit inertial-range scaling analogous to Kolmogorov theory, or
whether their behaviour is entirely dictated by intermittency. More
fundamentally, it is not known whether a single intermittency-based framework
suffices to describe the scaling of gradient observables.

In this work, we show that two-point velocity-gradient correlations obey a previously unrecognized parity-dependent scaling structure. Using exact relations and direct numerical simulations, we demonstrate that the second-order gradient correlation is fixed by the Laplacian of the velocity correlation and therefore scales as $r^{-4/3}$ in the inertial range. Strikingly, all odd--odd higher-order correlations inherit approximately the same scaling with remarkably weak dependence on order, whereas even--even correlations display systematically different exponents.

We show that this dichotomy originates from the sign structure of the gradient field. Rapid sign decorrelation suppresses intermittent contributions in odd--odd correlations, causing their leading behaviour to reduce to the velocity-controlled second-order scaling. In contrast, even--even correlations remain sensitive to intermittent structures and reflect their sparse spatial organization~\cite{Mukherjeeetal2024}. Their measured exponents are quantitatively consistent, across two Reynolds numbers, with independently measured box-counting dimensions of intense gradient regions.

These results identify parity under sign reversal as a fundamental organizing principle for turbulent observables beyond intermittency and establish an explicit connection between sparse intermittent geometry and scaling exponents of turbulent correlation functions.

\begin{figure}
	\includegraphics[width=1.0\linewidth]{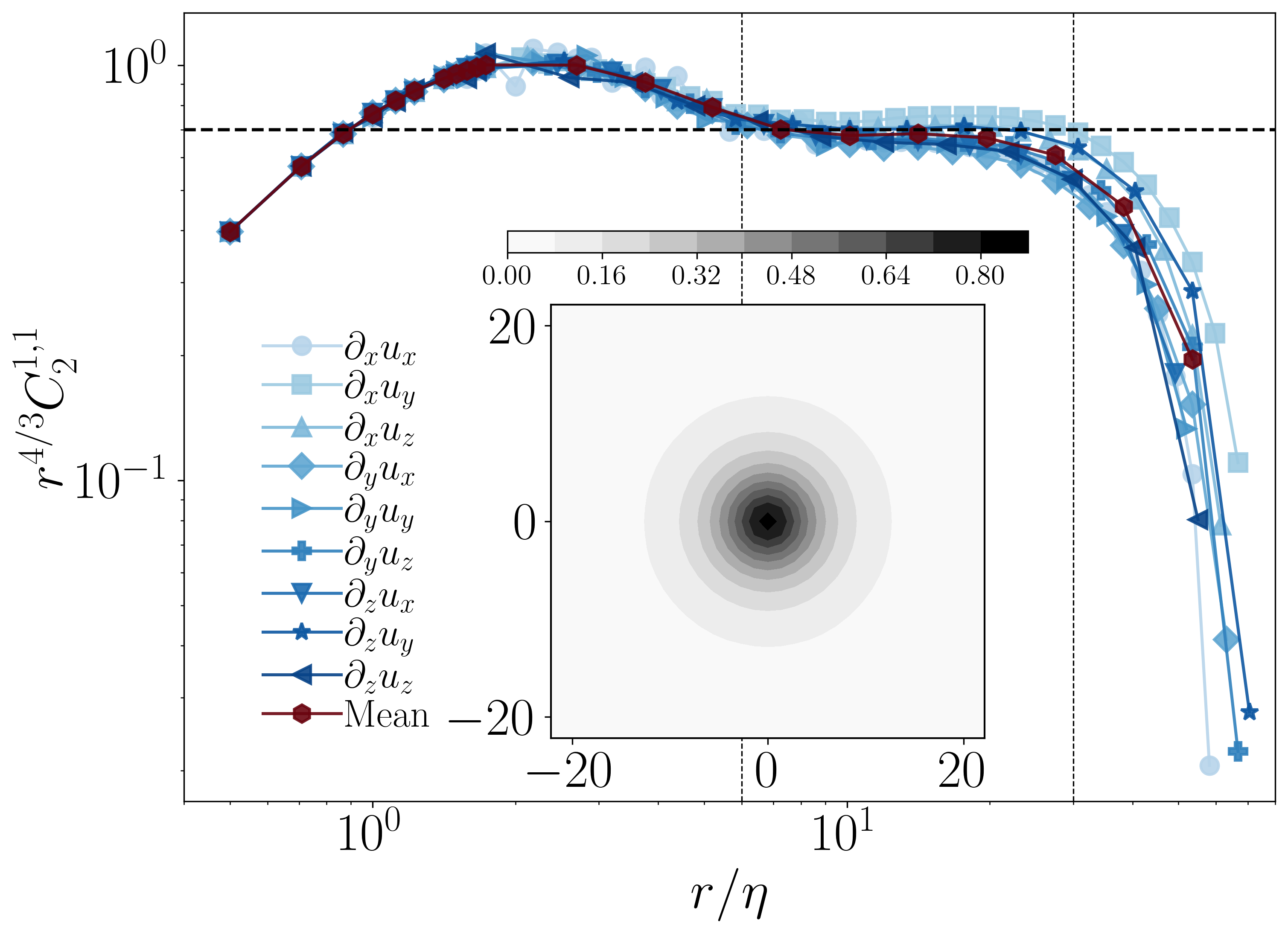}
	\caption{The compensated second-order gradient correlation $r^{4/3}C^{1,1}_2(r)$, 
	for different components and the mean, as a function of the normalised separation
	$r/\eta$ in log--log scale for ${\rm Re}_\lambda = 220$. The dashed horizontal line is a guide to the eye 
	to indicate the quality of scaling and agreement with the theoretical prediction. 
	Inset: Representative two-dimensional slice of the correlation
function $C_2^{1,1}(\mathbf{r})$, showing approximate isotropy
	with no preferred directional dependence from simulations with ${\rm Re}_\lambda = 350$.}
\label{fig:C2}
\end{figure}

\begin{figure*}
	\centering
\includegraphics[width=0.48\linewidth]{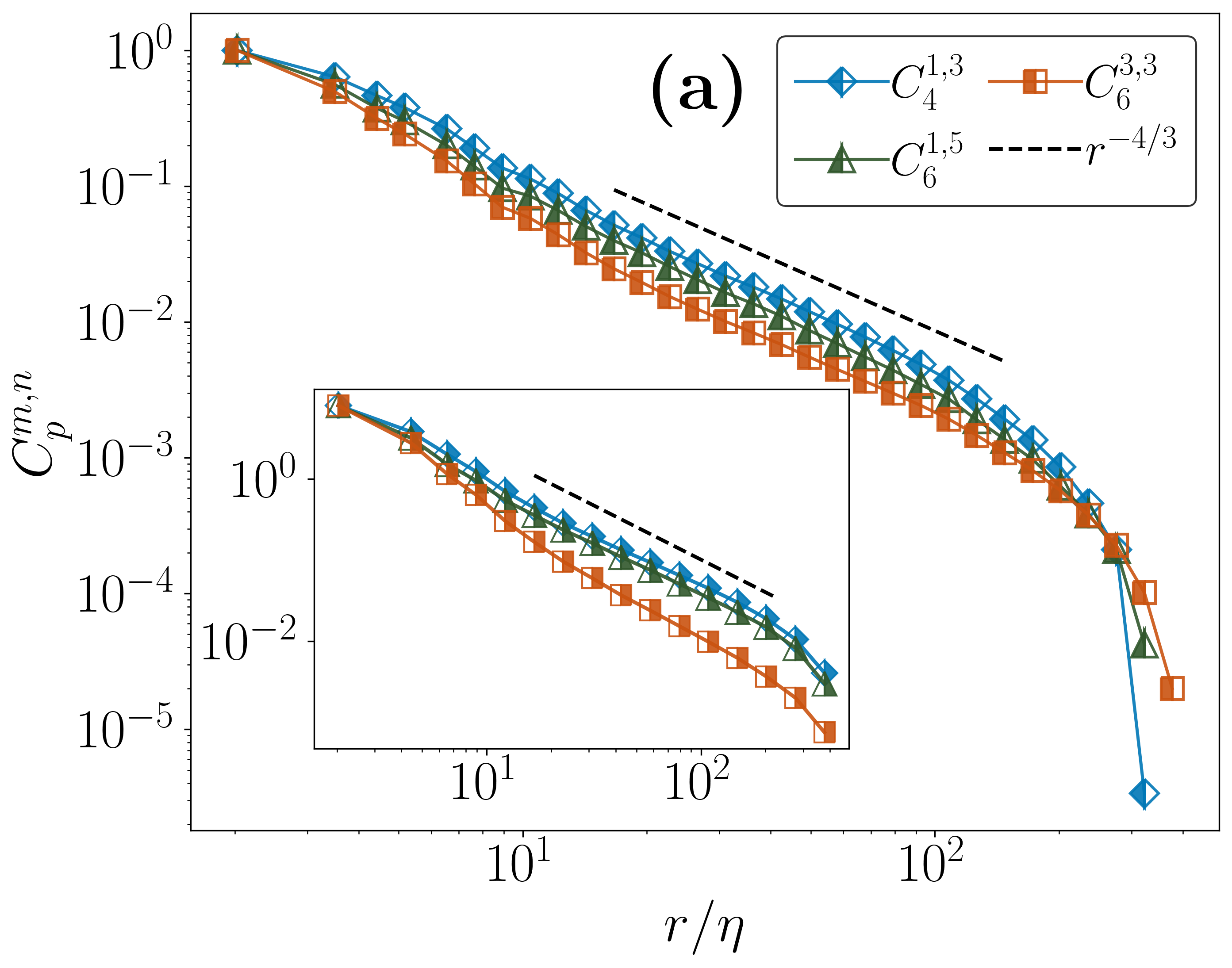}
\includegraphics[width=0.48\linewidth]{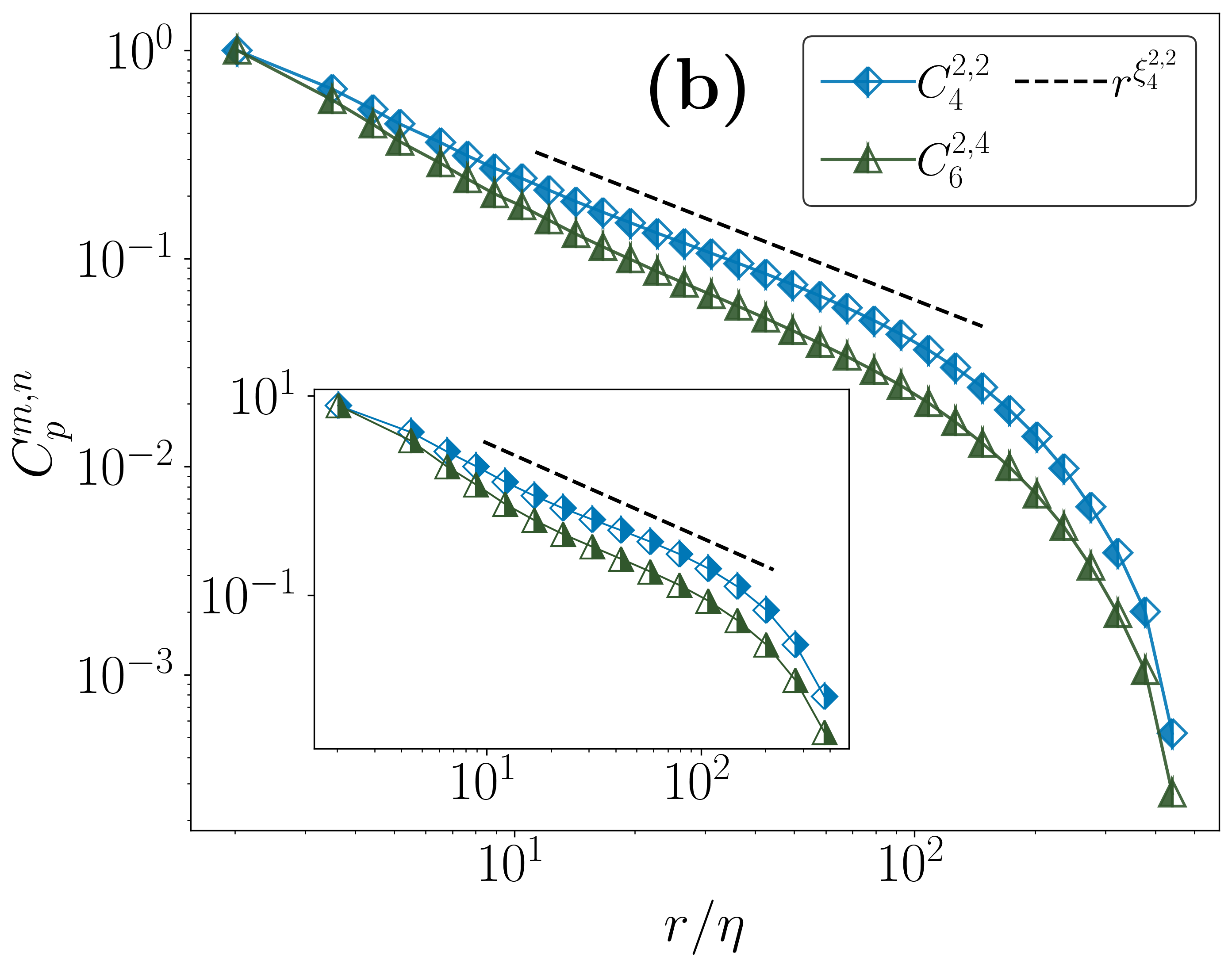}
	\caption{Higher-order gradient correlations $C_p^{m,n}(r)$ for different $(m,n)$ combinations for ${\rm Re}_\lambda = 220$: (a) odd--odd correlations, showing scaling close to $r^{-4/3}$; (b) even--even correlations, exhibiting systematically shallower exponents  $\xi_p^{m,n} > -4/3$. Insets: corresponding off-diagonal correlations.}
\label{fig:Cp}
\end{figure*}


Before turning to spatial correlations, we briefly recall the statistical
structure of the velocity-gradient tensor at a single point. In homogeneous
isotropic turbulence, the components of $\partial_j u_i$ are constrained by
incompressibility and rotational invariance, leading to nontrivial relations
between different components and between the symmetric (strain-rate) and
antisymmetric (rotation-rate) parts of the tensor~\cite{meneveau2011}. These constraints are well
satisfied in our numerical data and reflect the rich internal structure of the
velocity-gradient field. As we show below, however, this structure does not
affect the scaling of two-point gradient correlations, whose behaviour is
instead fixed by kinematic relations and inertial-range velocity statistics.

We define a representative component of the velocity-gradient tensor
as $g(\mathbf{x}) = \partial_j u_i(\mathbf{x})$, and consider the two-point
correlation
\begin{equation}
	C_p^{m,n}(\mathbf{r}) = \langle g^m(\mathbf{x})\, g^n(\mathbf{x+r}) \rangle,
\label{eq:Cmn}
\end{equation}
with $m,n \in \mathbb{N}$, where $\mathbb{N}$ is a positive integer, and total order $p=m+n$. 
Under isotropy, the correlation depends only on $r = |\mathbf{r}|$, so we write
$C_p^{m,n}(r)$.

We first examine the second-order case $p=2$ with $m=n=1$. Using
statistical homogeneity and the commutation of spatial derivatives with
ensemble averaging, derivatives with respect to $\mathbf{x}$ can be
transferred to derivatives with respect to the separation $\mathbf{r}$,
yielding~\cite{Batchelor,Pope}
\begin{equation}
C_2^{1,1}(r) =
-\nabla_{r}^2
\langle u_i(\mathbf{x})u_i(\mathbf{x}+\mathbf r)\rangle .
\label{eq:C2_laplacian}
\end{equation}
This relation fixes the scaling of the second-order gradient correlation
entirely in terms of inertial-range velocity statistics.  Expressing the velocity correlation in terms of the
second-order structure function $S_2(r)\sim r^{\zeta_2}$ in the inertial
range, we obtain
$C_2^{1,1}(r)\sim r^{\zeta_2-2}$.
Within Kolmogorov phenomenology $\zeta_2=2/3$, leading to
	$C_2^{1,1}(r)\sim r^{-4/3}$ and thus, for the velocity-gradient correlator in the inertial range 
	Eq.~\eqref{eq:C2_laplacian}, the second-order scaling exponent $\xi_2^{1,1} \equiv \zeta_2 - 2 \approx -4/3$. 

Figure~\ref{fig:C2} shows the two-point velocity-gradient correlation
$C_2^{1,1}(r)$ measured in direct numerical simulations as
a function of $r/\eta$. Data for different components and
Reynolds numbers collapse onto a common inertial-range
scaling consistent with the prediction
$C_2^{1,1}(r)\sim r^{-4/3}$ (dashed line), confirming that
the scaling is fixed by the relation in Eq.~(\ref{eq:C2_laplacian}).
The inset
shows a representative two-dimensional slice of
$C_2^{1,1}(\mathbf{r})$, demonstrating the approximate
isotropy of the measured correlation function.

Motivated by this result, we consider the scaling ansatz
\begin{equation}
C_p^{m,n}(r)\sim r^{\xi_p^{m,n}},
\end{equation}
for higher-order gradient correlations. Unlike the second-order case,
no exact relation analogous to Eq.~(\ref{eq:C2_laplacian}) constrains the
exponents $\xi_p^{m,n}$ for $p>2$, and their behaviour must therefore be
determined empirically. As we show below, the resulting scaling exhibits a
qualitative dependence on the structure of the correlations, with distinct
behaviour for different parity classes --- $m$ and $n$ both being even or both being odd --- and does not directly reflect the
intermittency of the underlying gradient field.


To investigate gradient correlation statistics, we perform direct
numerical simulations of the incompressible Navier--Stokes equations
in a triply periodic domain of size $L=2\pi$ using a pseudo-spectral
method~\cite{Canuto, Orszag1971}. Turbulence is maintained in a statistically stationary state
through a constant power injection~\cite{Pope_forcing} at a resolution $N=512^3$
yielding $Re_\lambda \approx 220$~\cite{Ray2018}. Additionally, we analyze data from the Johns Hopkins
Turbulence Database (JHTDB)~\cite{yeung2012dissipation,li2008public,perlman2007data}
at $N=1024$ ($Re_\lambda \approx 350$). Velocity fields are sampled over
many large-eddy turnover times to construct two-point gradient
correlation functions with high statistical accuracy~\cite{SHIRIAN2023106046}.

In constructing higher-order correlations, we subtract the mean from each
factor so that the correlations represent fluctuations at the prescribed order.
Without this subtraction, contributions from lower-order moments dominate at
large separations, leading to a flattening of the correlations and obscuring
inertial-range scaling. The scaling exponents reported in
Table~\ref{tab:exponents} are obtained by either computing the
exponent for each snapshot and component, followed by averaging over time and
over components (diagonal and off-diagonal treated separately) or from fits on
the time-averaged data; the quoted uncertainties correspond to the standard
deviation of these measurements in the first case and the fitting uncertainty
in the latter. The inferred exponents remain stable under moderate variations
of the fitting interval within the inertial range and across different gradient
components, indicating that the observed parity-dependent separation is not
driven by a particular choice of fitting window.

\begin{figure}
	\includegraphics[width=1.0\linewidth]{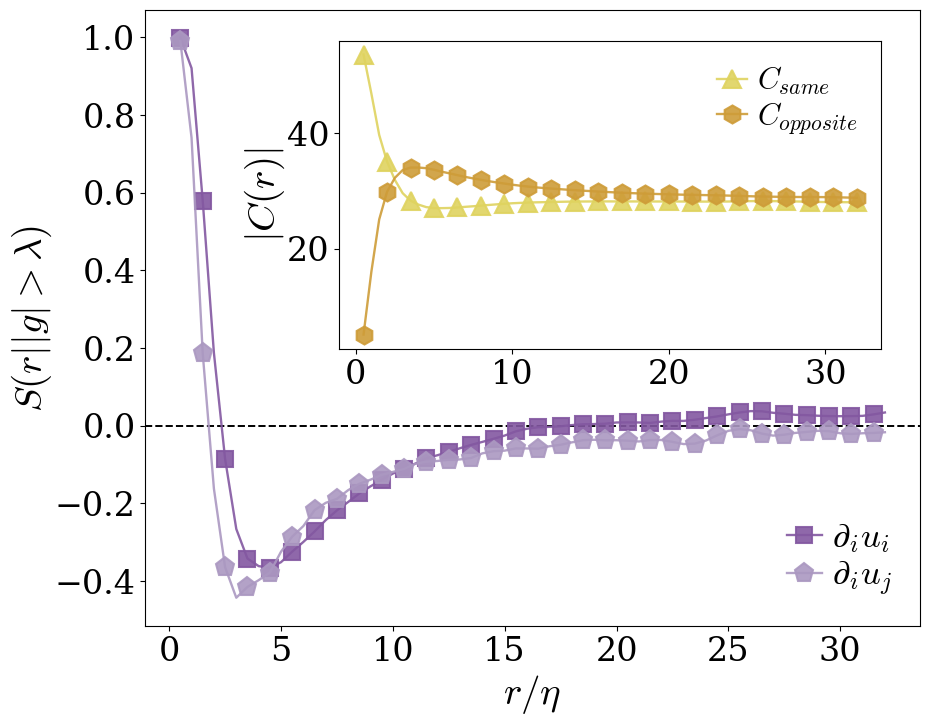}
\caption{Sign correlation $S(r)=\langle \mathrm{sign}(g(x))\,\mathrm{sign}(g(x+r))\rangle$, conditioned on $|g|>\lambda$, as a function of separation $r$,  showing weak scale dependence and values close to zero. Inset: decomposition of the gradient correlation into same-sign and opposite-sign contributions, demonstrating strong cancellation.}
\label{fig:cancel}
\end{figure}


\begin{figure*}
\includegraphics[width=1\linewidth]{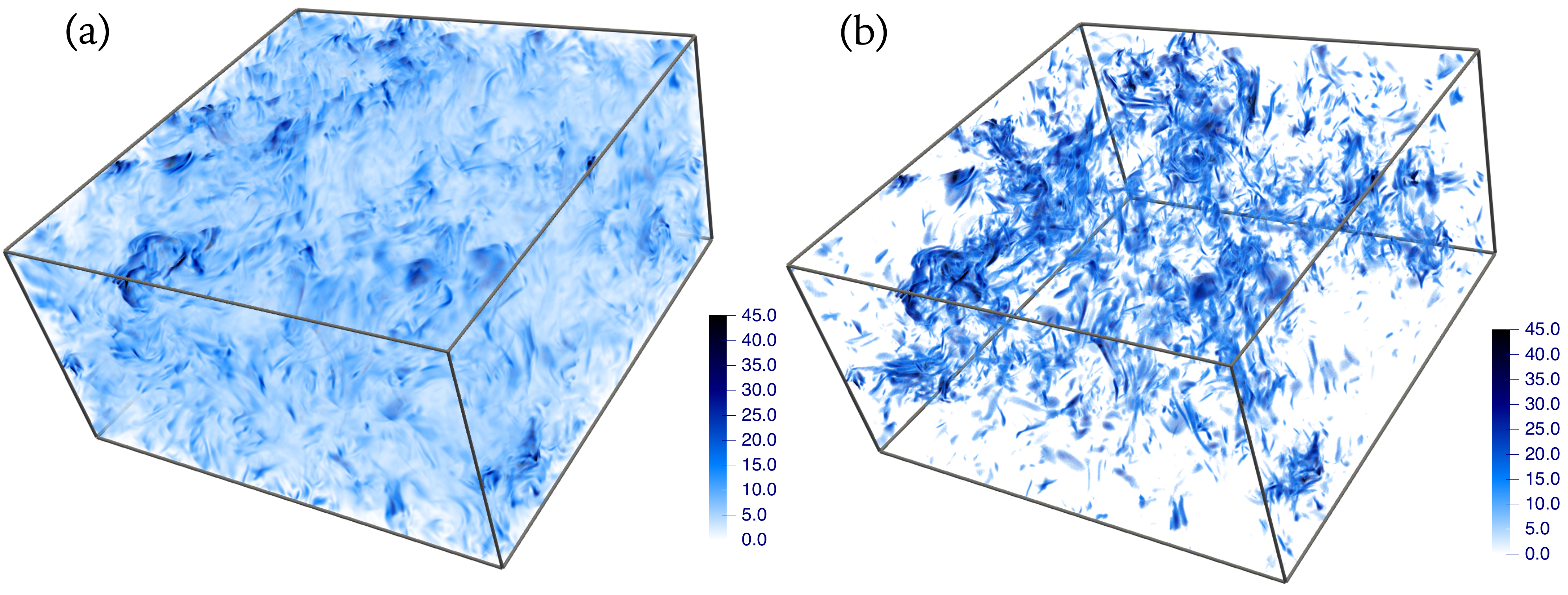}
	\caption{Three-dimensional visualization of the velocity-gradient field (for ${\rm Re}_\lambda = 350$):  
	(a) Magnitude of $|\partial_x u_x|$ and (b) regions satisfying the threshold condition $|\partial_x u_x|>\lambda\sigma$ with $\lambda=4$. The thresholded field reveals sparse, spatially clustered intermittent structures.}
\label{fig:Slice}
\end{figure*}


Figure~\ref{fig:Cp} shows representative examples of $C_p^{m,n}(r)$ for different $(m,n)$ combinations, grouped into (a) odd--odd and (b) even--even cases. Data from the $512^3$ and $1024^3$ simulations collapse well over the inertial range, and different gradient components, including the off-diagonal cases shown in the insets, exhibit similar behaviour. We restrict attention to correlations with $m$ and $n$ of the same parity. Mixed-parity cases (e.g., $p=3,5$) combine contributions with different sign structure and do not exhibit a clean inertial-range scaling, consistent with the competing sign contributions discussed later.

A clear distinction emerges between the two classes. Odd--odd correlations
exhibit an approximate power-law scaling close to $r^{-4/3}$, with only weak
dependence on order. In contrast, even--even correlations display
systematically different exponents that deviate from this value and vary  with
$(m,n)$. These trends are quantified in Table~\ref{tab:exponents}, which lists
the measured scaling exponents $\xi_p^{m,n}$.  Although the accessible inertial
range is necessarily limited at the present Reynolds numbers, the
parity-dependent organization remains robust across both datasets and across
diagonal and off-diagonal components.  Odd--odd exponents are consistent with
$-4/3$ within uncertainty and remain stable across both datasets. Even--even
correlations, in contrast, vary with $(m,n)$ and show a mild Reynolds-number
dependence, consistent with their sensitivity to the spatial organization of
intermittent structures through an effective dimension
$D(\lambda)$~\cite{Sreenivasan1991,SreenivasanAntonia1997} as  discussed later.

The near collapse of odd--odd correlations onto a common scaling, despite
increasing order, is striking: in contrast to conventional intermittency
phenomenology, higher-order correlations do not exhibit progressively stronger
deviations from dimensional scaling. The behaviour of even--even correlations
shows that this suppression is not universal, and indicates that different
classes of correlations probe distinct aspects of the gradient field.

Interestingly, the exponent $-4/3$ has also appeared in recent analyses of
coarse-grained velocity gradients~\cite{Eyink2007} based on inertial-range spectral
arguments~\cite{JohnsonWilczekReview2024}. However, those approaches concern
filtered second-order quantities, whereas here the same scaling re-emerges for
odd--odd higher-order correlations through a qualitatively different mechanism
involving parity-dependent suppression of intermittent contributions.

These observations suggest that the scaling of $C_p^{m,n}(r)$ is governed not
only by intermittency but also by the sign structure of the correlations
themselves. In particular, the weak dependence of odd--odd correlations on
order indicates that their dominant contributions originate from mechanisms
qualitatively different from those expected solely from intermittent
fluctuations. We now show that this behaviour arises from
a near cancellation between same-sign and opposite-sign intermittent
contributions, leading to strong suppression of odd--odd intermittent
correlations.

\begin{figure}
\includegraphics[width=1.0\linewidth]{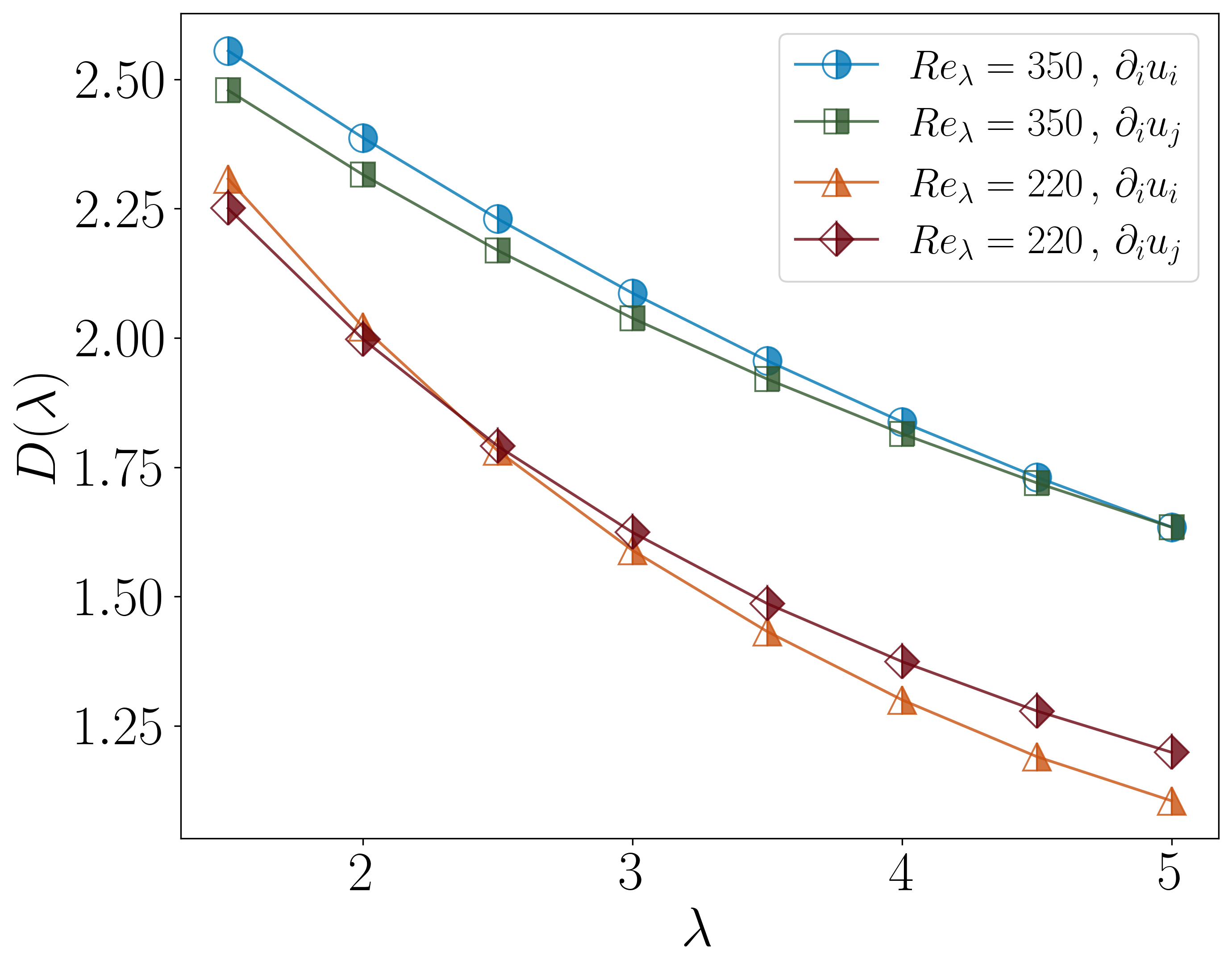}
	\caption{Box-counting dimension $D(\lambda)$ of intense velocity-gradient structures as a function of threshold $\lambda$ for the two Reynolds numbers considered in this work, showing progressively sparser spatial support with increasing threshold.}
\label{fig:D}
\end{figure}


To identify the dominant contributions, we decompose the field as
$g(\mathbf{x}) = b(\mathbf{x}) + I(\mathbf{x})$,
where $b$ represents a smooth background contribution and $I$ denotes sparse intermittent events. 
Expanding
\begin{equation}
C_p^{m,n}(r) = \langle b^m b^n \rangle + \langle I^m I^n \rangle + \text{mixed terms}.
	\label{eq:exp}
\end{equation}

Writing
$I(\mathbf{x}) = s(\mathbf{x}) |I(\mathbf{x})|$ with $s=\pm1$,
odd--odd correlations depend on the sign product
$s(\mathbf{x})s(\mathbf{x+r})$, whereas even--even correlations are invariant under sign reversal.
For odd $m,n$, the intermittent contribution reads
\begin{equation}
\langle I^m(\mathbf{x}) I^n(\mathbf{x+r}) \rangle
=
\langle s(\mathbf{x}) s(\mathbf{x+r})
\, |I(\mathbf{x})|^m |I(\mathbf{x+r})|^n \rangle .
\end{equation}
The inset of Fig.~\ref{fig:cancel} shows that the total correlation arises from a near cancellation between same-sign and opposite-sign contributions, which are individually large but opposite in sign. Figure~\ref{fig:cancel} shows that the sign correlation
$\langle s(\mathbf{x}) s(\mathbf{x+r}) \rangle$
remains small across the inertial range. We also observe approximate factorization,
$\langle s(\mathbf{x}) s(\mathbf{x+r}) |I(\mathbf{x})|^m |I(\mathbf{x+r})|^n \rangle
\simeq
\langle s(\mathbf{x}) s(\mathbf{x+r}) \rangle
\langle |I(\mathbf{x})|^m |I(\mathbf{x+r})|^n \rangle$,
consistent with strong suppression of intermittent odd--odd contributions.

While some of the mixed terms are suppressed because of a similar sign
decorrelation, a subset still survives.  As we show in the Appendix, the
surviving terms up to leading order have the same scaling form as the
correlation of the $b$ field.  Thus, the leading scaling behaviour of
$C_p^{m,n}(r)$, for odd $m$ and $n$, is controlled primarily by the background
term in Eq.~\eqref{eq:exp}. It is easy to show (Appendix) that a Wick-type factorization~\cite{Isserlis1918}
results in $\langle b^m b^n \rangle \sim \langle b ({\bf x}+{\bf r}) b({\bf x}) \rangle$.
Hence, the inertial-range behaviour is dominated by ordinary two-point
background correlations, which are fixed by the second-order result.
Consequently, $\xi_p^{m,n} \approx \xi_2^{1,1} \approx -4/3$, with only weak
dependence on $(m,n)$.  This is
consistent with the collapse observed for odd--odd correlations in
Fig.~\ref{fig:Cp} and the exponents reported in Table~\ref{tab:exponents}.

In contrast, even--even correlations are invariant under sign reversal and
therefore retain contributions from intermittent structures. To quantify their
spatial organization, we threshold the gradient field according to
$|g|>\lambda\sigma$, as illustrated in the three-dimensional visualization of
Fig.~\ref{fig:Slice}.  The resulting intense structures are sparse and strongly
clustered in space, with box-counting dimension $D(\lambda)$ decreasing
systematically with increasing $\lambda$, as shown in
Fig.~5~\cite{Sreenivasan1991,SreenivasanMeneveau1986}.  The probability that
two points separated by $r$ both belong to such regions therefore scales as
$P(r)\sim r^{D(\lambda)-3}$.
Hence, $\langle I^m(\mathbf{x}) I^n(\mathbf{x+r}) \rangle
\sim P(r)$.

The measured exponents, for the even--even sector, suggest that the background
term in Eq.~\eqref{eq:exp} is subdominant: $\xi_p^{m,n} > \xi_2^{1,1}$, for $m$
and $n$ even (see Table~\ref{tab:exponents}).  The mixed terms in this case are
subdominant, similar to the odd-odd sector (see Appendix).  Thus, for the
even--even sector $C_p^{m,n}(r) \sim \langle I^m(\mathbf{x}) I^n(\mathbf{x+r})
\rangle$ and hence $\xi_p^{m,n} \sim D(\lambda)-3 \neq \xi_2^{1,1}$.

The measured exponents are consistent with this picture. For the even--even
cases in Table~\ref{tab:exponents}, the observed $\xi_p^{m,n}\approx -0.8$ to
$-1.2$ correspond to $D (\lambda) \approx 1.8$--$2.2$ which lie within the
measured range of $D(\lambda)$ shown in Fig.~\ref{fig:D} for both Reynolds
numbers. In particular, the observed exponents are broadly consistent with the
dimensions obtained near $\lambda\approx2$, although at present we do not have
a theoretical argument fixing a preferred threshold value.  Nevertheless, the
numerical evidence $\xi_p^{m,n} > \xi_2^{1,1}$ constraints the effective
dimension to be greater than 5/3 and hence provides an (Reynold number
dependent) upper bound on $\lambda$ (see Fig.~\ref{fig:D}).  The agreement
between independently measured even--even exponents and the geometrical scaling
inferred from the box-counting analysis provides strong support for the
proposed mechanism.

These results provide a unified picture: odd--odd correlations suppress
intermittent contributions through sign decorrelation and reduce, to leading
order, to velocity-controlled scaling fixed by the second-order correlation,
whereas even--even correlations retain intermittent contributions and reflect
their spatial organization. Parity thus defines two distinct classes of
observables in turbulence beyond intermittency.


To summarise, we have investigated two-point velocity-gradient correlation
functions in homogeneous isotropic turbulence using exact relations and direct
numerical simulations. The second-order correlation obeys the exact relation
$C_2^{1,1}(r) = -\nabla_r^2 \langle u(x)u(x+r)\rangle$, which directly implies
inertial-range scaling $C_2^{1,1}(r)\sim r^{-4/3}$. 

At higher orders, we uncover a qualitatively new parity-dependent organization
of gradient correlations. Odd--odd correlations exhibit scaling close to
$r^{-4/3}$ with remarkably weak dependence on order, whereas even--even
correlations display systematically different exponents. We show that this
distinction originates from the sign structure of the gradient field: sign
decorrelation suppresses intermittent contributions in odd--odd sectors,
causing their leading behaviour to reduce to the velocity-controlled
second-order scaling, while even--even correlations retain intermittent
contributions and remain sensitive to the sparse spatial organization of
intense structures.

The measured even--even exponents are quantitatively consistent, across both
Reynolds numbers, with independently measured box-counting dimensions of
intermittent gradient structures. More broadly, these results suggest that
parity and sign structure can organize turbulent observables in ways not
captured by intermittency alone, establishing an explicit connection between
sparse intermittent geometry and scaling exponents in turbulence.

\begin{acknowledgements}
AD acknowledges Mrinal Jyoti Powdel and Anikat Kankaria for productive discussions. 
RM acknowledges the Infosys-TIFR Leading Edge Travel Grant 2025 which facilitated 
part of this research. AB acknowledges the Long Term Visiting Students’ Program (LTSVP) 
of the ICTS-TIFR which enabled this collaboration. SSR acknowledges the Indo–French Centre for the
Promotion of Advanced Scientific Research (IFCPAR/CEFIPRA, project no. 6704-1)
for support.  This research was supported in part by the International Centre
for Theoretical Sciences (ICTS) for the program --- 11th Indian Statistical
Physics Community Meeting (code: ICTS/11thISPCM2026/04). The simulations were
performed on the ICTS clusters Mario, Tetris, and Contra. AD, RM and SSR
acknowledge the support of the DAE, Government of India, under projects nos.
12-R\&D-TFR-5.10-1100 and RTI4001. 
\end{acknowledgements}

\bibliographystyle{apsrev4-2} 
\bibliography{references}

\appendix
\onecolumngrid

\section*{Appendix}

We briefly outline the mechanism underlying the parity-dependent scaling of the
higher-order gradient correlations discussed in the main text. We decompose the
velocity-gradient field as $g(x)=b(x)+I(x)$, where $b(x)$ denotes a
comparatively smooth background contribution and $I(x)=s(x)A(x)$ represents
sparse intermittent events, with $s(x)=\pm1$ and $A(x)\ge0$. Expanding the
correlation function, $C_p^{m,n}(r)=\langle g^m(x)g^n(x+r)\rangle = \langle b^m b^n \rangle + \langle I^m I^n \rangle + \text{mixed terms}$, generically
produces three classes of contributions: background terms $C_{bb}$ involving only $b$,
intermittent terms $C_{II}$ involving only $I$, and mixed terms containing both fields.
The behaviour of these contributions is governed by two empirical observations:
(i) rapid sign decorrelation, $\langle s(x)s(x+r)\rangle\approx0$ throughout
the inertial range; and (ii) sparse intermittent support characterized by a
pair probability $P(r)\sim r^{D(\lambda)-3}$. Together, these ingredients lead
to qualitatively different scaling behaviour for odd--odd and even--even
correlations.

\subsection*{Odd--Odd Correlations}

For odd $m$ and $n$, the purely intermittent contribution is
$I^m(x)I^n(x+r)=s(x)s(x+r)A^m(x)A^n(x+r)$.
Since the sign correlation $\langle s(x)s(x+r)\rangle$ remains small throughout the inertial range, 
the intermittent contribution $C_{II}(r)=\langle I^m(x)I^n(x+r)\rangle$
is strongly suppressed. Numerically, we further observe approximate factorization,
$\langle s(x)s(x+r)A^m(x)A^n(x+r)\rangle\approx\langle s(x)s(x+r)\rangle\langle A^m(x)A^n(x+r)\rangle$
consistent with near cancellation between same-sign and opposite-sign contributions.

We next consider the background contribution
$C_{bb}(r)=\langle b^m(x)b^n(x+r)\rangle$. 
The background field $b(x)$ is comparatively smooth and weakly intermittent.
We therefore assume that higher-order connected cumulants of the background
field are subleading in the inertial range, so that the dominant separation
dependence of $\langle b^m(x)b^n(x+r)\rangle$ is captured primarily by
Wick-type factorization into disconnected pairwise contributions~\cite{Isserlis1918}.
Schematically, contractions among fields at the same point contribute only
local moments, while contractions spanning the separation $r$ carry the
inertial-range dependence. For odd $m$ and $n$, the leading contribution
contains the smallest possible number of cross-separation contractions. For
example,
$\langle b^3(x)b^3(x+r)\rangle = 9\,\langle b^2(x)\rangle \langle b^2(x+r)\rangle \langle b(x)b(x+r)\rangle +
6\,\langle b(x)b(x+r)\rangle^3 +\cdots$, 
where the ellipsis denotes subleading connected contributions. The first term
contains the smallest possible number of cross-separation contractions and
therefore dominates in the inertial range, while terms involving higher powers
of $\langle b(x)b(x+r)\rangle$ decay more rapidly with $r$. The dominant
inertial-range dependence is therefore inherited from a single two-point
correlation factor. Since $\langle b(x)b(x+r)\rangle\sim r^{-4/3}$, the
background contribution inherits the same leading inertial-range scaling,
$C_{bb}(r)\sim r^{-4/3}$.

Finally, we consider the mixed terms. Contributions containing an odd number of
intermittent factors inherit a sign factor $s(x)$ or $s(x+r)$ and therefore
undergo the same cancellation as the purely intermittent sector. The remaining
mixed terms contain only even powers of $I$ and therefore survive sign
cancellation. However, these terms require intermittent events at both points
and hence carry the sparse geometric factor $P(r)\sim r^{D(\lambda)-3}$.  Since
the surviving mixed contributions still contain at least one background
cross-correlation, their leading behaviour scales schematically as $C_{\rm
mix}(r)\sim r^{D(\lambda)-3}r^{-4/3}$ which decays faster than the background
contribution alone.  The dominant inertial-range behaviour is therefore
governed by the background sector, $C_p^{m,n}(r)\sim r^{-4/3},$ independently
of the order $m+n$, and hence $\xi_p^{m,n}\approx \xi_2^{1,1}$.

\subsection*{Even--Even Correlations}

The even--even sector is qualitatively different because even powers are
invariant under sign reversal. We first consider the behaviour expected if the
correlations were governed solely by the background field $b(x)$.

Assuming, as in the odd--odd sector, that higher-order connected cumulants of
the background field are subleading, the dominant contributions are again
captured by Wick-type pairwise contractions~\cite{Isserlis1918}. For example,
$\langle b^2(x)b^2(x+r)\rangle = \langle b^2(x)\rangle \langle b^2(x+r)\rangle
+ 2\langle b(x)b(x+r)\rangle^2$.  The first term is independent of $r$, while
the leading decaying contribution scales as $\langle b(x)b(x+r)\rangle^2\sim
r^{-8/3}$.  More generally, higher even--even moments involve progressively
higher powers of $\langle b(x)b(x+r)\rangle$, leading to increasingly steep
inertial-range decay. A weakly connected background field therefore predicts
rapidly decaying even--even correlations together with constant offsets.

This is not what is observed numerically. The measured even--even exponents are
substantially shallower and therefore cannot be explained through pairwise
background correlations.

For even powers, $I^{2m}(x)I^{2n}(x+r)=A^{2m}(x)A^{2n}(x+r)$, so the
sign-cancellation mechanism responsible for suppressing odd--odd intermittent
contributions no longer operates. Expanding again gives
$C_p^{2m,2n}(r)=C_{bb}(r)+C_{II}(r)+C_{\rm mix}(r)$.

Intermittent events occupy only a sparse fraction of the flow volume, with pair
probability
$P(r)\sim r^{D(\lambda)-3}$.
The intermittent contribution therefore scales as
$C_{II}(r)\sim \langle A^{2m}(x)A^{2n}(x+r)\rangle P(r)$. 
Since the intermittent amplitudes vary only weakly with separation across the
inertial range, the dominant $r$-dependence of $C_{II}(r)$ is inherited
primarily from the geometric factor $P(r)$.

Unlike the odd--odd sector, mixed terms in the even--even case do not undergo
sign cancellation and therefore generically survive. These terms also require
intermittent events at both points and hence carry the same geometric factor
$P(r)\sim r^{D(\lambda)-3}$.
Schematically,
$C_{\rm mix}(r)\sim P(r)b^kA^\ell$, 
whereas
$C_{II}(r)\sim P(r)A^{m+n}$, 
with $\ell<m+n$. Both contributions therefore inherit the same geometric
scaling exponent $D(\lambda)-3$, while differing in their amplitudes through
the number of intermittent factors involved. Since intermittent amplitudes
satisfy $A\gg b$, the purely intermittent contribution is expected to dominate
numerically, although the inertial-range scaling itself is governed more
generally by the sparse geometry of intermittent structures which appear in both the terms,

Thus, $C_p^{2m,2n}(r)\sim P(r) \sim r^{D(\lambda)-3}$.  Since
$D(\lambda)-3>-8/3$ over the measured range of thresholds, the intermittent
contribution decays substantially more slowly than the background contribution
and therefore dominates the inertial-range scaling. The resulting exponents
$\xi_p^{m,n}\approx D(\lambda)-3$ are thus controlled primarily by the spatial
organization of intermittent structures rather than by pairwise background
correlations, explaining both the comparatively shallow scaling exponents and
their strong parity dependence relative to the odd--odd sector.

\begin{table*}
\centering
\begin{tabular}{|c|c|c||c|c||c|c|}
\hline
\multicolumn{3}{|c||}{} & \multicolumn{2}{c||}{$Re_\lambda = 220$} & \multicolumn{2}{c|}{$Re_\lambda = 350$} \\
\hline
\hline
	& \multicolumn{2}{|c||}{} 
	& Diagonal & Off Diagonal 
	& Diagonal & Off Diagonal \\
	\hline
	p & m & n  
	& $\xi_p^{m,n}$ & $\xi_p^{m,n}$  
	& $\xi_p^{m,n}$  & $\xi_p^{m,n}$ \\
\hline
	& \multicolumn{2}{|c||}{Odd} 
	&  &  
	&  &  \\
	\hline
2 & 1 & 1 &  $-1.37 \pm 0.08$  &  $-1.37 \pm 0.08$ & $-1.318 \pm 0.006$ & $-1.250 \pm 0.007$ \\
\hline
4 & 1 & 3 & $-1.35 \pm 0.09$ &  $-1.4 \pm 0.1$  & $-1.346 \pm 0.005$ & $-1.312 \pm 0.006$ \\
\hline
6 & 1 & 5 & $-1.4 \pm 0.2$ & $ -1.5 \pm 0.3$  & $-1.403 \pm 0.007$ & $-1.353 \pm 0.009$ \\
\hline
6 & 3 & 3 & $-1.5 \pm 0.1$ & $-1.8 \pm 0.4$ & $-1.402 \pm 0.006$ & $-1.569 \pm 0.008$ \\
\hline
	& \multicolumn{2}{|c||}{Even} 
	&  &  
	&  &  \\
	\hline
4 & 2 & 2 & $-1.0 \pm 0.1$ & $-1.1 \pm 0.1$ & $-0.765 \pm 0.004$ & $-0.824 \pm 0.003$ \\
\hline
6 & 2 & 4 & $-1.1 \pm 0.1$ & $-1.2 \pm 0.1$ & $-0.870 \pm 0.004$ & $-0.929 \pm 0.003$ \\
\hline
\hline
\end{tabular}
\caption{Scaling exponents for gradient correlations $C_p^{m,n}(r)$ for different $(m,n)$ combinations, 
reported separately for diagonal and off-diagonal components with estimated uncertainties. Odd--odd cases are close to $-4/3$, 
while even--even cases are systematically shallower.}
\label{tab:exponents}
\end{table*}

\end {document}